\documentclass[10pt,conference,letterpaper]{IEEEtran}
\usepackage{times,amsmath,epsfig}

\usepackage{type1cm}     
\usepackage{algorithm}     
\usepackage{algorithmic}     
\usepackage{graphicx}     
\usepackage{epstopdf}
\usepackage{tikz}     
\usetikzlibrary{arrows.meta,positioning,shapes.geometric,fit,backgrounds,calc}
\definecolor{figink}{HTML}{3A3A3A}
\definecolor{figgray}{HTML}{EEEEEE}
\definecolor{figaccent}{HTML}{2F6DB5}
\definecolor{figaccentlight}{HTML}{DCE8F6}
\tikzset{
  fig/.style={font=\sffamily\footnotesize, line width=0.6pt, draw=figink, text=figink,
    >={Stealth[length=5pt,width=4pt]}},
  box/.style={draw=figink, fill=figgray, rounded corners=2pt, minimum height=6mm,
    minimum width=16mm, align=center, inner sep=3pt},
  store/.style={draw=figaccent, fill=figaccentlight, rounded corners=2pt, minimum height=6mm,
    minimum width=16mm, align=center, inner sep=3pt},
  db/.style={cylinder, shape border rotate=90, aspect=0.3, draw=figaccent, fill=figaccentlight,
    align=center, inner sep=3pt},
  op/.style={circle, draw=figink, fill=figgray, minimum size=11mm, align=center, inner sep=1pt},
  group/.style={draw=figink!60, dashed, rounded corners=3pt, inner sep=4pt},
  flow/.style={->, draw=figink},
  biflow/.style={<->, draw=figink},
  lbl/.style={font=\sffamily\scriptsize, text=figink, align=center},
  title/.style={font=\sffamily\footnotesize\bfseries, text=figink},
}
\newcommand{\stackedop}[3]{
  \node[op] (#1c) at ($(#2)+(-0.14,0.14)$) {};
  \node[op] (#1b) at ($(#2)+(-0.07,0.07)$) {};
  \node[op] (#1) at (#2) {#3};
}

\usepackage{xspace}     
\usepackage{booktabs}     
\usepackage[bf,tableposition=top]{caption}     
\usepackage[hyphens]{url}     
\usepackage[bookmarks, pdftex, colorlinks=false]{hyperref}     
\usepackage[square,numbers]{natbib}     
\usepackage[font=small,labelfont=scriptsize]{caption}
\usepackage[section]{placeins}

\newcommand{\spara}[1]{\smallskip\noindent{\bf #1}}

\usepackage[show]{chato-notes}
\usepackage{amsmath,amssymb}

\newcommand{\tuple}[1]{\ensuremath{\langle #1 \rangle}\xspace}

\title{Fault Tolerance for Stream Processing Engines}

\author{
{
Muhammad Anis Uddin Nasir
}
\vspace{1.6mm}\\
\fontsize{10}{10}\selectfont\itshape
KTH Royal Institute of Technology, Stockholm, Sweden\\
\fontsize{9}{9}\selectfont\ttfamily\upshape

anisu@kth.se
}

\begin{document}
\pagestyle{headings}  

\maketitle

\begin{abstract}
Distributed Stream Processing Engines (DSPEs) target applications related to continuous computation, online machine learning and real-time query processing.
DSPEs operate on high volume of data by applying lightweight operations on real-time and continuous streams.
Such systems require clusters of hundreds of machines for their deployment.
Streaming applications come with various requirements, i.e., low-latency, high throughput, scalability and high availability. 
In this survey, we study the fault tolerance problem for DSPEs. 
We discuss fault tolerance techniques that are used in stream processing engines that were state of the art at the time of writing (2016), namely S4, Storm, Flink, Spark Streaming, Samza and MillWheel.
Further, we give insight on fault tolerance approaches that we categorize as active replication, passive replication and upstream backup.
Finally, we discuss implications of the fault tolerance techniques for different streaming application requirements.
This revised version corrects errors in the original 2016 manuscript, summarizes later developments in Flink and Spark, discusses durable logs and archival storage for replay, and points to more recent surveys.
\end{abstract}

\maketitle
\section{Introduction}
\label{sec:intro}
Distributed Stream Processing Engines (DSPEs) handle continuous stream of data and guarantee real-time processing with high scalability \cite{neumeyer2010s4,toshniwal2014storm,abadi2005design,shah2003flux,zaharia2012discretized}.
DSPEs target applications like online machine learning, continuous computation, real-time query processing and others. 
For example, fraud detection, network monitoring, anomaly detection, credit-card authorization, advertising services, financial markets, and user behavior monitoring are some of the common streaming applications.
Streaming applications deal with high volume of data that is generated by sensor networks, stock markets, call data records (CDRs), web traffic, etc \cite{gedik2014partitioning}. 

Clusters with hundreds of machines are used for practical and scalable deployment of streaming frameworks at companies like Google and Yahoo. 
However, as the number of machines in a cluster grow, it increases the likelihood of failure of a single machine inside a cluster at any given moment.
Failure of machines includes node failures, network failures, software bugs and resource limitations \cite{marz2015big}.
Moreover, streaming applications run for indefinite period, which increases the chances for disruption due to unexpected failures or planned reboots.
Failures in DSPEs may cause an application to block or produce erroneous results. 
As a consequence, along with scalability and low latency requirements, many critical streaming applications require DSPEs to be highly available \cite{stonebraker20058}.

\begin{figure}[t]
\begin{center}
\begin{tikzpicture}[fig]
  \coordinate (s) at (0,0); \coordinate (w) at (3,0);
  \stackedop{src}{s}{Source}
  \stackedop{wrk}{w}{Worker}
  \node[op] (agg) at (6,0) {Aggre-\\gator};
  \draw[flow] (src) -- node[lbl, above] {key\\grouping} (wrkc.west |- wrk.west);
  \draw[flow] (wrk) -- node[lbl, above] {key\\grouping} (agg);
  \draw[flow] ($(srcc.west)+(-0.9,0)$) -- node[lbl, above] {input\\stream} (srcc.west);
\end{tikzpicture}
\caption{An example of a multi-stage DAG that consists of input stream, multiple streaming operators and  data channels. Input continuously arrives in the form of stream, operators perform transformation on the incoming stream and forward it to other streams using the data channels. }
\label{fig:dag}
\end{center}
\end{figure}

Stream processing frameworks differ from other distributed systems, e.g., batch processing, cloud-based and database systems, due to the requirement of real time data processing.
DSPEs operate on unbounded streams, rather than bounded data. 
Thus, such systems cannot easily predict the trends for incoming set of events, e.g., data distribution, concept drifts, incoming rate, etc. 
DSPEs require low-latency processing on real-time data.
Therefore, it is very common in DSPEs to apply light weight operations, like filters, aggregators, and union, and to keep expensive input/output operations, like writing to and reading from external storage, off the critical processing path.
Yet, strong processing guarantees require some information to be persisted, which creates a tension between low latency and fault tolerance.
Lastly, due to the aforementioned requirements, fault-tolerance techniques used in other distributed systems cannot always be directly applied to DSPEs.
In contrast, DSPEs require fault tolerant techniques with very low recovery time to support time-critical, real-time and continuous applications.

For last decade, batch processing is used to implement various big data applications \cite{zaharia2012resilient, hindman2011mesos, vavilapalli2013apache}.
Specially, MapReduce \cite{dean2008mapreduce}, a programming model that has been adopted for implementation of batch processing applications on large datasets. 
Such systems execute algorithms on large amount of stored data and require significant amount of time for data processing. 
However, as discussed above, there are various applications that require real-time computation, and cannot be served using traditional batch processing, due to their high volume and low latency requirement.
There have been efforts in the past making batch processing systems work for streaming setting \cite{condie2010mapreduce,neumeyer2010s4}.
Nevertheless, such tweaks are inefficient for stream processing.

Streaming Applications are often represented using a directed acyclic graph (DAG), where vertices are the operators and the edges are the channels for dataflow between operators.
Figure \ref{fig:dag} shows an example DAG for a streaming application.
For example, the same DAG can be used to implement a na\"{i}ve Bayes classifier \cite{murphy2006naive}.
The incoming stream is forwarded to the first level of operators in a round robin manner.
Further, downstream operators route each feature and class value further downstream, where frequency of each feature and class value is aggregated to calculate the final aggregate of each feature.

In order to implement fault tolerant DSPEs, partial information like internal states of operators and routing decisions must be replicated, so that the information can be reconstructed in case of a failure. 
In case of DSPEs, various techniques have been proposed for fault tolerance that are categorized as, active replication \cite{shah2004highly,balazinska2008fault,ananthanarayanan2013photon,loesing2012stormy}, passive replication \cite{castro2013integrating,kwon2008fault,brito2009minimizing,sebepou2011cec,zhou2010programming} and upstream backup \cite{qian2013timestream,akidau2013millwheel}.
In active replication, input stream is forwarded to multiple processing nodes, which process same sequence of events and coordinate together to produce a consistent output.
In passive replication, nodes are organized in master/slave or primary/secondary order, where primary processes the input and sends the snapshots (checkpoints) to secondary nodes periodically for backup. 
In upstream backup, upstream operators keep track of the tuples that they forward downstream until they are acknowledged.
Combinations of these techniques are also commonly used to achieve fault tolerance, e.g., SGuard \cite{kwon2008fault} and the scheme of Fernandez et al. \cite{castro2013integrating} combine passive replication with upstream backup to provide fault tolerance for stateful operators. 
Active operator replication incurs high cost by creating multiple copies of each operator, whereas passive replication leads to high recovery time, and upstream backup is inefficient for recovering the complete history of an operator \cite{hwang2005high}. 

Fault tolerance techniques vary among applications, on the basis of recovery models. 
Recovery methods have been categorized into precise, rollback and gap recovery by \cite{hwang2005high}. 
In precise recovery, no information loss is guaranteed, but it increases the latency and reduces the throughput of the overall system. 
Precise recovery is implemented by modifying the passive replication technique to track each tuple.
In rollback recovery, system tries to recover from the footprints of the previous execution and induces lesser latency compared to precise recovery.
In case of node failure, nodes recover its state using the last checkpoints and the tuples from the upstream output queue.
However, in case of timeout, upstream nodes replay the event, hence, the system might process some of the tuples more than once. 
Lastly, in gap recovery, the information loss is expected with minimum effect on latency on the system.

DSPEs provide different reliability models for various applications, i.e., at-most-once, at-least-once and exactly-once semantics.
These models vary among each other based on the number of times a tuple is allowed to be processed.
Choice of fault tolerance scheme is directly related with the reliability requirement of set of applications that will be running on a DSPE.

\spara{Note on this revision.}
This version corrects errors in the background, system descriptions and cost analysis of the original 2016 manuscript.
The descriptions of stream processing engines reflect the state of the art at the time of writing (2016).
Since then, the field has evolved considerably, e.g., consistent distributed snapshots and exactly-once state have become standard in engines such as Apache Flink \cite{carbone2015lightweight}.
For Flink and Spark, which remain widely used open-source DSPEs, we briefly summarize the main developments since 2016 in Section \ref{sec:stream}.
We also discuss the role of durable logs, such as Kafka, and of archival object storage in fault tolerance (Section \ref{subsec:logs}).
For more recent treatments of state management and fault tolerance in stream processing, we refer the reader to the surveys of To et al. \cite{to2018survey}, Wang et al. \cite{wang2022comprehensive} and Fragkoulis et al. \cite{fragkoulis2024survey}.


\section{Background}
\label{sec:background}
This section introduces different concepts and methodologies related to DSPEs. In general, to support DSPEs, new data models and operators have been proposed, which differ from traditional DBMS and batch processing systems.
\subsection{Stream Processing Models}
Streaming applications are commonly represented by a directed acyclic graphs (DAG). 
They take an input stream, which is unbounded sequence of tuples \tuple{k,v} where $k$ is the key and $v$ is the value and process the stream by passing through the DAG.
A streaming DAG contains set of vertices and edges.

The vertices in a DAG, often called a processing element (PE), represent streaming operators that apply transformations on the incoming streams.
Throughout this survey, we use PE and operator interchangeably to represent a streaming operator.
Each operator in a DAG contains their input and output queue. 
PEs process the data by applying certain operations like, filters, union, aggregators, and joins, and may produce some output.

The edges in a DAG represent the flow of stream between operators (see Figure \ref{fig:dag}).
Output messages are stored in the output queue before the delivery to a downstream operators.
Fault tolerance for streaming application requires making components of a DAG resilient towards failures.
These components include both vertices (operators)  and edges (channels).
In particular, operators can implement the fault tolerance for these components using their input and output queue, i.e., keep tuples in input queue until they are processed and store tuples in output queue until they are delivered.

\subsection{Streaming Operations}
Most of the DSPE operations are lightweight and are implemented in-memory to achieve low latency processing for streaming applications.
As DSPEs are inspired from traditional DBMS, they provide implementation of various relational operators like select, join, project, union and aggregate. 
To understand various operations, we categorize these operators in groups based on their sensitivity. \cite{brito2009minimizing}.

\subsubsection{Stateless}
These operators do not maintain or require any extra information or local state for their operation.
One of the simplest examples of a stateless streaming operator is filtering.
Filtering is analogous to select operator in DBMS, where operators discard or forward tuples based only on their attributes. 
Such operators are deterministic and stateless, as they do not have any time or order dependency and do not require any local state for their functionality.
Such operators are not affected by any out-of-order or delayed arrival of input stream. 
Fault tolerance for stateless operators is achieved by being able to replay the stream. 
This is implemented by upstream operator maintaining the stream until the downstream operators process and acknowledge them to the upstream operator. 
In case of failure, re-spawning the failed operator and processing the queued messages will provide fault tolerance for stateless operators.

\subsubsection{Stateful and Deterministic}
These operators require input tuple along with current state of the system to apply transformation.
A deterministic operator does not have any time dependency on its input streams.
Few common examples of stateful and deterministic operations are calculating aggregates, combining multiple events based on count and time window.
Such operators implement functions such as sum, max, min, average and others.

For instance, to implement time window aggregation, each operator requires to maintain a local partial count for all the unique tuples that it has observed during the time window.
In case of failure of stateful deterministic components, it is necessary either to store the local state of a PE or to replay all the events to reconstruct the state.
A good approach to implement fault tolerance for stateful and deterministic operators is to execute periodic checkpoints along with upstream backup (see Section \ref{sec:approach}). 
Upstream operators require keeping track of the tuples after the last checkpoint. 
In case of a failure, both checkpoints and upstream buffer are used to recover the state of the operator.
This approach reduces the size of the log and speeds up the recovery process, as it does not require replaying the complete sequence of past events.
\subsubsection{Stateful and non-Deterministic}
These operators require the local state of an operator and their output also depends on factors that are not fixed by the input, such as the relative arrival order of tuples across multiple input streams, the processing (wall-clock) time, or randomness.
Consequently, replaying the same input after a failure may not reproduce the same state or output.
Operators that consume multiple streams are common examples, e.g., joins in DBMS.
For example, a stream S1 can be joined with stream S2 using a common attribute such as primary key in databases.
Such operators deal with multiple streams to produce a new event with a combined operation.
A join becomes non-deterministic when its result depends on the order in which tuples from the different input streams are interleaved, rather than only on their contents and timestamps.
Operators that order their inputs deterministically, e.g., by event timestamps, can be made deterministic.
Further, a union operator that merges multiple streams into a single stream in arrival order is a common example of stateful and non-deterministic operator.
Also, operators that aggregate based on processing-time windows follow a similar pattern. 
All these operators require more expensive approaches to achieve fault tolerance.
A simple fault tolerance approach for stateful and non-deterministic operator is to perform a complete checkpoint, before forwarding any tuple downstream.
\subsection{Fault Models}
\subsubsection{Crash-Stop}
This model represents the common modes of failure that exist in modern DSPEs, such as node hangs and crashes.
In this model, a worker never recovers after a crash. 
On failure, a worker loses its local state, its locally persisted state and all the in-flight messages.
This model allows nodes to fail at any time.
In case of failure, a failed node stops communicating with other nodes in the system.
Other remaining nodes can detect such failures using heartbeat messages, failure detectors or keep-alive messages.
Fault tolerance in the crash-stop model is implemented using replication (see Section \ref{sec:approach}).
\subsubsection{Omission Failure}
Another kind of fault in distributed systems can be omission faults. 
Such faults occur when a process does not send or receive a message. 
It may be caused due to buffer overflows or network congestion. 
With an omission, a worker drops messages that are supposed to be exchanged with other processes.
\subsubsection{Crash-Recovery}
In this model, a worker fails, but might recover after some time. 
This model resembles practical deployments, where failed workers are restarted, e.g., by a cluster manager.
A correct worker may crash, but recovers in finite time and eventually stays up.
A worker that crashes and never recovers or crash infinitely many times (unstable worker) are said to be incorrect or faulty workers.
This scheme is often implemented using a stable storage that maintains the states of workers, which can be used to reconstruct the states in case of failure. 
Crash-Recovery model is expensive compared to Crash-stop model, as it involves the expensive storage step.
\subsection{Reliability Models}
Heterogeneous message processing guarantees are required for stream processing applications.
For example, an approximation algorithm could afford to lose a message, whereas, a financial transaction could not.
Therefore, DSPEs provide various reliability models for different applications.
\subsubsection{At-most-once}
This mode guarantees that each message is processed at maximum once. 
At-most-once processing allows messages to be dropped in case timeout or failure.
It ensures that duplicated tuples are not processed more than once.
This mode does not require any special handling, i.e., in case of a failure of a worker, restarting the worker or spawning a new worker serves the requirement.
\subsubsection{At-least-once}
This mode replays each message in case of timeout or failure. 
At-least-once model requires tracking each message upstream, while it gets processed downstream. 
It does not require any implementation to handle duplicated tuples.
In case of failed processing (due to timeout or failure), messages are replayed.
Therefore, each message may get processed more than once.
\subsubsection{Exactly-once}
This model ensures that the effect of each tuple on the state and output of the application is reflected exactly once, even if the tuple is physically processed more than once after a failure.
Therefore, it requires replaying the failed tuples and needs to avoid applying the effects of duplicated messages.
It is typically implemented by combining consistent checkpoints of the operator state with replay of the input after the last checkpoint, together with deduplication, or with idempotent or transactional writes to the output. 
Exactly-once semantics are required for critical applications, like financial transactions.
\subsection{Recovery Models}
Hwang et al. \cite{hwang2005high} categorize recovery guarantees into a) gap recovery, b) roll-back recovery and c) precise recovery, based on different requirements of streaming applications.
\subsubsection{Gap Recovery}
\label{ref:gap}
This scheme follows fail-over or fail-safe mechanism, i.e., workers are allowed to fail and new worker takes over failed worker and start processing new tuples.
Therefore, there is state loss expected in Gap recovery.
Gap recovery provides weakest recovery guarantees among all the three recovery models and operates on most recent available information.
This strategy provides at-most-once message processing guarantees and targets stateless streaming applications (e.g., sensor-based environment monitoring).
\subsubsection{Roll-back Recovery}
It avoids message loss through replication.
This scheme combines checkpointing with upstream backup for recovery of failed workers.
In case of failure of workers, new worker takes over the failed worker by recovering its state from the last checkpoint.
As tuples after the last checkpoint are replayed, the output may contain duplicate tuples.
Thus, rollback recovery is appropriate for applications that requires no information loss, but can tolerate approximated or imprecise output, e.g., fire alarms, theft prevention through asset tracking.
This scheme guarantees at-least-once processing for each tuple and can be modified to achieve strict processing guarantees like precise recovery. 
\subsubsection{Precise Recovery}
This scheme provides the strongest recovery guarantees and hides the effects of a failure perfectly. 
It is the most expensive of the three recovery schemes in terms of recovery time.
Precise recovery modifies the recovery phase of roll-back recovery to guarantee that there is no information loss.
It ensures that the new worker starts exactly from the point where failed worker stopped. 
Precise recovery guarantees exactly-once semantic for the tuples and is well suited for financial services with strict correctness requirements.

\subsection{Requirements \& Challenges}
\label{subsec:rnc}
In this subsection, we discuss challenges that a DSPE requires to consider while choosing the most suitable fault tolerance scheme as per requirement.
\subsubsection{Latency}
As DSPEs target applications like continuous computation and real-time processing, low latency is one of the most important characteristics. Streaming applications achieve low latency by processing messages on the fly, by performing lightweight operations and by avoiding the costly storage operations in the critical processing path \cite{stonebraker20058}.
Most DSPEs keep their state in memory and avoid synchronous writes to disk or external storage on the critical processing path, as such writes incur high I/O cost. 
Fault tolerance requires replication, and induces an additional overhead on a streaming application. 
Therefore, there is a strong need of considering application requirements while choosing a proper fault tolerance technique.
\subsubsection{Stream Ordering}
Unlike traditional batch processing and DBMS systems, DSPE deals with real-time and continuous data streams, which may contain delayed, missing, or out-of-sequence data events.
These delayed events can be caused for various reasons, such as, variable sources, network delays, and loss events.
To handle the variance in the input streams, DSPEs require intelligent fault tolerance schemes that are capable of segregating between unordered and failed tuples.
\subsubsection{State Management}
As discussed in previous sections, fault tolerance requires replicating the information.
Replication cost varies based on the fault tolerance scheme selected, e.g., active replication requires at least 2x storage cost, whereas in passive replication scheme, additional storage is required for checkpointing, and upstream backup requires additional memory to keep unprocessed tuples in its output queue.
In case of passive replication, the runtime overhead is proportional to the frequency of checkpointing.
Additionally, fault tolerance requires out-of-the-box implementation for state management, i.e., checkpointing, backup, restore and partitioning \cite{castro2013integrating}.
\subsubsection{Recovery Time}
For time-critical streaming applications, such as radar monitoring, traffic control monitoring, financial transactions and fraud detection systems, recovery time is one of the critical factors.
Such systems require strict bounds on the latency and throughput of streaming applications.
Hence, fault tolerance scheme requires such considerations to meet desired SLAs.

\section{Fault Tolerance Approaches}
\label{sec:approach}
Fault tolerance in distributed systems has been extensively studied \cite{cristian1991understanding}.
Most of the schemes rely on replication by creating multiple copies of the information in the system, and recovers from the copies in case of failures \cite{gray1996dangers}.
We categorize the fault tolerance approaches for stream processing systems into three categories that are a) active replication, b) passive replication and c) upstream backup. 
Few solutions use combination of these approaches.
\subsection{Active Replication}
Active replication replicates operators and processes each input tuple on multiple operator instances. 
It provides an efficient solution for fault tolerance with low recovery time at the cost of replication. 
Active replication involves two expensive steps that are ordered multicast and output synchronization.
Ordered multicast ensures that all nodes receive the input in same order.
Whereas, output synchronization ensures delivery of correct and single output to the client.
Most of the information processing using active replication are redundant and require at least twice more storage and processing resources.
Therefore, it is an expensive fault tolerance scheme; and is feasible for geo-replicated DSPEs. 
Active replication does not require system to store any state in stable storage.

Flux \cite{shah2003flux}, a dataflow operator that is inserted between operators (producer-consumer) to perform adaptive state partitioning and dataflow routing for streaming applications.
It targets the load balancing and fault tolerance problem for deterministic stateful operators.
It solves the load balancing problem by adaptively repartitioning of the state of operators during execution.
Flux uses process-pair approach \cite{shah2004highly}, to achieve fault tolerance. 
It uses redundant copies of dataflow computation, which allows quick fail-over and automatically recovers without affecting the execution.
Recovery process in flux is divided into two phases, i.e., \textit{take over} and \textit{catch up}.
In case of failure, a replica machine \textit{takes over} the processing.
Once the failed machine recovers, the \textit{catch up} phase start and both machines (replica and recovered machine) start processing the tuples.
Flux guarantees exactly-once semantics for streaming application and provides in-order delivery of in flight tuples by attaching a sequence number. 

Borealis \cite{abadi2005design} uses DPC \cite{balazinska2008fault} (Delay, Process, and Correct), a protocol to handle crash failures in a DSPE.
It deals with deterministic operators and provides fault tolerance by replicating all operators on multiple processing nodes.
Each operator is replicated, and nodes switch to replica in case of failure of upstream operators.
To ensure replica consistency, Borealis uses a SUnion data serializing operator.
DPC offers configurable trade-off between availability and consistency.
In particular, it allows users to define their desired bounds for availability and attempts to minimize the inconsistencies between replicas during partitions.
Moreover, it ensures eventual consistency and provides exactly-once semantics for the data processing.
In case of failure, Borealis delay the processing of tuples without violating the availability bound, while maintaining the output consistent.
Further, in case the recovery takes more time than the availability bounds, it compromises the consistency to satisfy the availability requirement.
However, it guarantees reconciliation of inconsistent results by re-running their computation on the correct input stream, after recovery.

Photon \cite{ananthanarayanan2013photon}, is a geographically distributed and fault tolerant system for joining multiple continuously flowing streams.
It is capable of handling unordered or delayed streams.
Photon provides at-most-once at all times and eventual exactly-once semantics for operators.
Photon achieves data-center level fault tolerance using a geo-replicated Paxos-based Identifier Registry (IdRegistry) service, which is geo-replicated to achieve fault tolerance.
The IdRegistry provides a service for verification of existing events.
Therefore, each worker requires writing every event to the IdRegistry service.

Stormy \cite{loesing2012stormy}, is a cloud based streaming service that uses successor list replication to process multiple copies of each event.
Multiple nodes handle each event, where one of the nodes is selected as master, which is responsible to forward and coordinate the event among other replicas.
The master node and its replica form a replication group.
Each incoming event, is received at a master node that perform two steps: a) processing the event and b) forwarding the events to all the nodes in the replication group.
The event is processed on each replica node on their arrival and responses are acknowledged back to  the master.
The master node wait for majority quorum to accept each event as a correct or failed one.

\subsection{Passive Replication}
Passive replication (also known as primary-backup replication) is a replication scheme in which nodes are arranged in master-slave or primary-secondary relationship.
Only one front-end (primary) server handles the client request.
Other nodes in the cluster act as a backup nodes, which are responsible for storing snapshots (typically called checkpoints) of the master node.
Backup nodes take over the primary node in case of failure and recover from the stored checkpoints. 
Different fault tolerance techniques have been proposed for DSPEs using passive replication scheme.
These scheme vary based on different checkpointing strategies (i.e., using distributed storage or transactional memory) and recovery methods (i.e., parallel recovery).
Commonly, passive replication is used along with upstream backup for precise recovery of an operator, i.e., keep the tuples after the checkpoint in the output buffer and use them during recovery phase.

Fernandez et al.~\cite{castro2013integrating} achieves fault tolerance for deterministic operators by externalizing internal operator states. 
The proposed scheme periodically checkpoints state of an operator and maintains unprocessed tuples, after the checkpoint, at an upstream operators. 
The checkpointing occurs in a asynchronous manner to avoid the overhead.
To recover a failed stateful operator, the upstream operator with the most recent checkpointed state of the failed operator requests a new operator and restores the failed operator from that checkpoint. 
Since the unprocessed tuples are buffered by the upstream operator, they are replayed for complete recovery after the checkpoint. 
To speed up operator recovery, the failed operator may be scaled out before recovery.

SGuard \cite{kwon2008fault} asynchronously checkpoints the deterministic operator states on periodic basis and recovers from the recent state in case of a failure. 
It checkpoints the states of operators into a distributed and replicated file system, rather than the memory.
This leaves the memory available for streaming operators. 
Along with distributed storage, it uses a scheduler and memory management middleware to isolate checkpointing process from the general stream processing.
Given a queue of write requests, scheduler selects the replicas for each data chunk and schedules all writes in a manner that significantly reduces the latency of individual writes, while only modestly increasing the completion time for all writes in the queue.
Management middleware enables asynchronous, fine-grained checkpointing without interfering with streaming operators.

ChronoStream \cite{7113328}  delivers fault tolerance for deterministic operators using asynchronous delta checkpointing.
It provides lightweight state management by slicing the application-level states into computational slices.
The set of computational slices defines a computational state of an operator, where each slice is a computationally-independent unit associated with the input stream.
ChronoStream stores every update to a computational state in the key-value store.
Further, it asynchronously checkpoints the updates in the key-value store that are used to reconstruct an operator state in case of failure.
The reconstruction process recovers lost slices in parallel.

Sebepou et al. proposed continuous eventual checkpointing (CEC) \cite{sebepou2011cec}, a light weight rollback recovery mechanism for window-based streaming operators.
CEC splits the operator state into independent components (similar to window-based operators) and provides fault-tolerance guarantees by taking continuous incremental state checkpoints.
It represents the checkpoints as control tuples that are intermixed with the regular tuples at the output of an operator.
The control tuples represent the state of the system that includes the state and the ordering of all the open windows. 
CEC requires the output to be written into a stable storage, which is used during the recovery phase.
During the recovery phase, CEC reconstruct the operator state by rolling back on the output queue log.

Brito et al. \cite{brito2009minimizing} propose a scheme for minimizing the latency of non-deterministic operators using speculative execution \cite{brito2008speculative}.
The proposed scheme improves the latency of non-deterministic operators by relying on speculation to parallelize stateful components, using software transactional memory (STM).
The STM is an optimistic concurrency control mechanism that speculatively executes the transactions in parallel by storing transient states of the operators.
Further, upon conflicts, it automatically aborts or restarts the transaction. 
This enables STM to perform optimistic parallelization for costly operations. 
STM enables the system to store states in parallel, therefore, increasing the overall throughput and minimizing the latency of a DSPE.

TimeStream \cite{qian2013timestream} provides fault tolerance for deterministic operators and guarantees exactly once semantics for streaming applications.
It relies on tracking fine-grained dependencies between input and output streams, which can be used to re-compute the state in case of failures.
In case of failure of an operator or a sub-DAG, TimeStream replaces the failed component by creating a new instance of the failed portion of the DAG on a different machine.
The replacement mechanism is named as resilient substitution that ensures the resilience to faults by reproducing the state of an operator using the stored dependencies and upstream backup.

\subsection{Upstream Backup}
In upstream backup (UB), nodes store tuples until the downstream nodes process and acknowledge them. 
This scheme is different than active and passive replication in the way it handles failures.
UB uses the output queues at different workers as a temporary storage to store information.
This scheme is complicated as multiple downstreams operators process a single tuple.
The upstream backup model is feasible for operators whose internal state depends on a small amount of input.
It is often used in combination with passive replication. 
In case of failure, the upstream primaries replay their logs and the secondary nodes rebuild the missing state.
When used along with passive replication, nodes can recover its state using the last checkpoints and the upstream tuples with timestamp after the last checkpoint. 
We have already discussed schemes that use a combination of upstream backup and passive replication.

\subsection{Durable Logs and Archival Storage}
\label{subsec:logs}
In practice, upstream backup is increasingly delegated to a durable, partitioned and replicated log, such as Apache Kafka \cite{kreps2011kafka}, that sits in front of the DSPE.
Instead of keeping tuples in the output queues of upstream operators, the sources of a streaming application record the offsets up to which they have consumed the log, typically as part of a checkpoint.
In case of failure, the application restores its state from the checkpoint and replays the log from the recorded offsets.
As the log is stored and replicated independently of the DSPE, the replay does not depend on the availability of upstream operators.
Moreover, the same log can be replayed by multiple applications, which also enables reprocessing of the stream, e.g., after a code change (see Section \ref{subsec:kappa}).

However, a log only retains records for a configured period of time or up to a configured size\footnote{\url{https://kafka.apache.org/documentation/\#topicconfigs_retention.ms}}, and log compaction retains only the latest record for each key.
Hence, the offsets recorded in a checkpoint must still be retained at the time of recovery.
If an application is down for longer than the retention period, or if it is restored from an old checkpoint or savepoint (see Section \ref{subsec:flink}), the records after those offsets may no longer be available, which degrades the recovery to gap recovery.
Similarly, reprocessing or backfilling the stream is limited to the retention window, while retaining the complete stream on the local disks of the log brokers is expensive.

Cloud-native object storage, such as Amazon S3 or Google Cloud Storage, offers a cheaper and highly durable tier for such data.
First, tiered storage\footnote{\url{https://cwiki.apache.org/confluence/display/KAFKA/KIP-405\%3A+Kafka+Tiered+Storage}} offloads older log segments to object storage while preserving their offsets, which extends the window from which a stream can be replayed.
Second, the stream can be continuously archived to object storage, e.g., as columnar files, to keep its complete history.
In case of a failure that exceeds the retention period, or for backfilling, an application can read the historical part of the stream from the archive and then switch to the live log, e.g., using the hybrid source of Flink\footnote{\url{https://cwiki.apache.org/confluence/display/FLINK/FLIP-150\%3A+Introduce+Hybrid+Source}}.
Such schemes extend upstream backup with an inexpensive durable tier, at the cost of a lower replay throughput from object storage and the need to align the position in the archive with the offsets in the log to avoid gaps or duplicates.
Object storage is also commonly used to store the checkpoints of passive replication.

\section{Alternative Models}
Fault tolerance in streaming environment is considered as an expensive operation.
Therefore, various frameworks provide weak fault tolerance guarantees for data processing, to accommodate for higher throughput.
There has been research in architectures where a stream processing engines is accepted to have a lossy nature. 
To avoid faults in those scenarios, we need to have some estimation mechanism that enables answering queries through approximation.
In this section, we discuss two different models that can be used to eliminate the errors that exist in streaming applications due to their lossy nature.
\subsection{Lambda Architecture}
Lambda Architecture (LA) \footnote{\url{http://lambda-architecture.net/}} is a form of generic and scalable form of active replication that combines batch processing and stream processing (see Figure \ref{fig:lambda}) \cite{marz2015big}.
The input is fed into both batch processing and stream processing systems in parallel.
The transformation logic is implemented twice, once in batch system and once in the stream processing system. 
LA requires additional serving layer to merge the output coming from both the batch and real-time layer, which may be used to answer incoming queries.

\begin{figure}[ht]
\begin{center}
\begin{tikzpicture}[fig]
  \node[store] (topic) at (0,0) {input\\topic};
  \node[group, fit=(topic), inner sep=5pt, label={[title]above:Kafka}] (kafka) {};
  \node[box] (speed) at (2.6,0.75) {processing\\job};
  \node[group, fit=(speed), inner sep=3pt, label={[title]above:Speed layer}] (sl) {};
  \node[box] (batch) at (2.6,-0.95) {processing\\job};
  \node[group, fit=(batch), inner sep=3pt, label={[title]below:Batch layer}] (bl) {};
  \node[db, minimum width=16mm, minimum height=22mm] (dbs) at (5.0,-0.1) {};
  \node[title] at ($(dbs.north)+(0,0.18)$) {Serving DB};
  \node[store, fill=white, minimum width=13mm, minimum height=4.5mm, font=\sffamily\scriptsize] (st) at (5.0,0.15) {speed table};
  \node[store, fill=white, minimum width=13mm, minimum height=4.5mm, font=\sffamily\scriptsize] (bt) at (5.0,-0.5) {batch table};
  \node[box, fill=white, minimum width=10mm] (app) at (7.3,-0.1) {App};
  \draw[flow] (kafka.east) -- (sl.west);
  \draw[flow] (kafka.east) -- (bl.west);
  \draw[flow] (sl.east) -- (st.west);
  \draw[flow] (bl.east) -- (bt.west);
  \draw[biflow] (dbs.east |- app) -- node[lbl, above=1pt] {queries} (app.west);
\end{tikzpicture}
\caption{Lambda Architecture with Kafka acting as a data source, a stream processing system, e.g., Storm, as the speed layer and a batch processing system, e.g., Hadoop, as the batch layer (based on \cite{marz2015big}).}
\label{fig:lambda}
\end{center}
\end{figure}

LA targets applications that are built around complex asynchronous transformations that need to run with low latency.
It provides the reprocessing functionality that is ignored by most of the stream processing systems.
For example, an update in code (bug fixes, additional features) requires recomputation of output to see the effect of the change. 
One of the major motivation of LA is that real-time processing involves approximation and is more lossy than batch processing.
However, one downside of LA is accurate code maintenance for both batch and stream processing system.

LA provides fault tolerance by applying the data transformation twice to the incoming data and later efficiently producing a single output.
SummingBird \cite{boykin2014summingbird}, an open-source domain-specific language implemented in Scala and designed to integrate online and batch MapReduce computations in a single framework.
It uses Hadoop for batch processing and Storm for online processing.
SummingBird transparently integrates batch processing and stream processing and produce consistent results over time.

\subsection{Kappa Architecture}
\label{subsec:kappa}
Kappa Architecture (KA) is somewhat similar to the Lambda Architecture \footnote{\url{http://radar.oreilly.com/2014/07/questioning-the-lambda-architecture.html}}.
However, these two differ in the sense that KA relies on using combination of multiple instance of stream processing systems, where as LA relies on batch processing system.
Unlike LA, KA performs recomputation on the incoming stream to produce the results. 
In particular, KA is designed to perform recomputation to process code changes.
However, this approach can easily be adopted to achieve fault tolerance through reprocessing of input stream.

Figure \ref{fig:kappa} shows an example of KA, where Kafka cluster is used as a data source.
Multiple instance of stream processing systems may run to process the incoming stream.
Further, results are merged at the serving database to produce consistent results, which are used to respond to client queries.

\begin{figure}[ht]
\begin{center}
\begin{tikzpicture}[fig]
  \node[store] (topic) at (0,0) {input\\topic};
  \node[group, fit=(topic), inner sep=5pt, label={[title]above:Kafka}] (kafka) {};
  \node[box] (j1) at (2.7,0.45) {job version $n$};
  \node[box] (j2) at (2.7,-0.45) {job version $n{+}1$};
  \node[group, fit=(j1)(j2), inner sep=4pt, label={[title]above:Stream processing}] (sp) {};
  \node[db, minimum width=16mm, minimum height=22mm] (dbs) at (5.15,-0.1) {};
  \node[title] at ($(dbs.north)+(0,0.18)$) {Serving DB};
  \node[store, fill=white, minimum width=13mm, minimum height=4.5mm, font=\sffamily\scriptsize] (t1) at (5.15,0.2) {table $n$};
  \node[store, fill=white, minimum width=13mm, minimum height=4.5mm, font=\sffamily\scriptsize] (t2) at (5.15,-0.45) {table $n{+}1$};
  \node[box, fill=white, minimum width=10mm] (app) at (7.25,-0.1) {App};
  \draw[flow] (kafka.east) -- (j1.west);
  \draw[flow] (kafka.east) -- (j2.west);
  \draw[flow] (j1.east) -- (t1.west);
  \draw[flow] (j2.east) -- (t2.west);
  \draw[biflow] (dbs.east |- app) -- node[lbl, above=1pt] {queries} (app.west);
\end{tikzpicture}
\caption{Kappa Architecture with Kafka acting as a data source. Multiple versions of a stream processing job process the incoming stream. Results are merged in the serving DB (based on \cite{marz2015big}).}
\label{fig:kappa}
\end{center}
\end{figure}
\section{Analysis}
\label{sec:analysis}
For analysis, we consider various requirements that are common in streaming applications, i.e., latency, storage cost and recovery time.
In particular, stateless operations do not require any replication.
So, we consider stateful operations for analysis.
We analyze basic fault tolerance techniques that are active replication, passive replication and upstream backup.

We begin our analysis by considering an input stream with a sequence of $m$ messages, arriving at a rate of $\lambda$ messages per time unit.
The arrival rate varies throughout the life span of an application, but, for analysis, we consider a constant arrival rate.
Also, we simplify our analysis by considering that the input stream arrives in order.
Each message carries a key drawn from a key universe $\mathcal{K}$ that contains $K = |\mathcal{K}|$ unique keys.
Further, we consider a set of $w$ workers (operator-level parallelism) that process the input stream.
Each worker maintains a state entry for each unique key that it receives.
In the worst case, e.g., when messages are not partitioned by key, every worker receives every key, so the total state of the operator is $wK$ entries.
For active replication, we denote the number of replicas of each worker by $R \geq 1$.
For passive replication, we denote the checkpointing interval by $T_c$ (i.e., the checkpointing frequency is $f = 1/T_c$).
For upstream backup, each upstream worker maintains an output queue of at most $n_o$ messages.
The summary of the notation is: $m$ (stream length), $\lambda$ (arrival rate), $K$ (unique keys), $w$ (workers), $R$ (replicas), $T_c$ (checkpointing interval) and $n_o$ (output queue size).
\subsection{Storage Cost}
Active replication processes the input stream on $R+1$ copies of each worker (one primary and $R$ replicas), i.e., if a single cluster requires 10 machines, then we need $10(R+1)$ machines.
Each copy maintains the complete state of its worker.
Therefore, the total memory required is $(R+1)wK$, i.e., at least twice the memory of a system without fault tolerance.
As stream processing systems deal with huge datasets with millions of unique keys, active replication is an expensive strategy in terms of storage cost.

On the other hand, passive replication requires checkpointing.
The workers require $wK$ memory to maintain their states.
Additionally, the system requires stable storage to keep the latest checkpoint of the state of each worker, which is $wK$ in the worst case (multiplied by the replication factor of the stable storage, if it is replicated).
Hence, passive replication requires no additional memory, and a single copy of the state in stable storage, which is typically far cheaper than the $RwK$ additional memory required by active replication.

Upstream backup requires additional memory for the output queues.
Therefore, the $w$ workers require $wK + wn_o$ memory.
However, the output queues must retain every message that is required to rebuild the state of a failed downstream worker.
Thus, $n_o$ is bounded only when the state of an operator depends on a bounded amount of recent input, e.g., a window of length $W$, for which the queues hold $wn_o \approx \lambda W$ messages in total.
For such operators, upstream backup requires the least amount of storage of the three approaches.
For operators whose state depends on the complete history of the stream, the queues grow with $m$, which makes upstream backup impractical on its own.

The combination of passive replication and upstream backup only requires the output queues to retain the messages that arrived after the last checkpoint, i.e., $wn_o \leq \lambda T_c$.
Hence, it requires $wK + \lambda T_c$ memory and $wK$ stable storage.

\subsection{Latency and Throughput}
Active replication requires two expensive steps that are ordered multicast and output synchronization.
The choice of these algorithms directly affect the latency of the streaming application.
Alongside, passive replication involves periodic checkpointing and the throughput of the application is dependent directly on the frequency of checkpointing.
In upstream backup, the throughput of the system is proportional to the size of the output queue.
There is a tradeoff between the memory and the throughput, i.e., increasing size of the output queue leaves less memory for general processing and reduces the throughput. 
Also, reducing the size of output queue reduces the throughput, as upstream operators require waiting for the pending inflight tuples to be acknowledged.
Lastly, the combination of passive replication and upstream backup reduces the dependency on the output queue, as output queue only require keeping the tuples after the last checkpoint, hence, increasing the throughput.
\subsection{Recovery Time}
Active replication requires the least recovery time, as application only requires changing the routing to another node.
Passive replication requires producing the state from the checkpoints. 
However, this step can be sped up using parallel recovery.
When combined with upstream backup, the recovery also replays the messages that arrived after the last checkpoint, i.e., at most $\lambda T_c$ messages, so a shorter checkpointing interval reduces the recovery time at the cost of more frequent checkpoints.
Upstream backup does not require any additional step and replaying the events from upstream operators recovers the lost tuples, while the recovery time grows with the number of messages that must be replayed.
\section{Stream Processing Engines}
\label{sec:stream}
\subsection{S4}
S4 \cite{neumeyer2010s4} is a scalable, pluggable and a partially fault-tolerant DSPE that was developed at Yahoo!.
The architecture of S4 resembles actors model \cite{agha1985actors} that allows massively concurrent implementation of applications.
In S4 the only way of interaction between PEs is through emission and consumption of data events.
Moreover, each PE is capable of instantiating new PEs, e.g., implementing the traditional wordcount example in S4 requires instantiating a new PE for each new word. 
Therefore, in applications with larger number of unique keys, there is a need of garbage collections of PE objects over time.
Each PE consumes events that correspond to it (e.g., hashing the key to select the PE), with the exception of keyless PEs (keyless PEs are typically used as input layers). 
The communication layer uses ZooKeeper \cite{zookeeper} and provides cluster management, automatic failover to standby nodes and maps physical nodes to logical nodes.

In S4 implementation, fault tolerance guarantees are provided using fail-over mechanism (Gap Recovery).
It uses ZooKeeper to manage all the alive nodes, and in case of failure, it automatically detects the failed node and routes all the messages to the standby node.
However, the local state of the failed node is lost while using the fail-over mechanism.
S4 also provides checkpointing and recovery support.
Still, developers need to inject code for checkpointing module and require providing backend storage implementation\footnote{\url{http://incubator.apache.org/s4/doc/0.6.0/fault_tolerance/}}.
\subsection{Storm}
Storm \cite{toshniwal2014storm} is a real-time, fault-tolerant and distributed stream data processing system.
Like other streaming frameworks, Storm represents applications as a DAG and call them topologies.
PEs in Storm are divided into two types, i.e., \textit{spout} and \textit{bolts}. 
Spouts are responsible to pull the data from external sources such as Kafka \cite{kreps2011kafka} and Kestrel\footnote{\url{http://robey.github.com/kestrel}}.
Bolts apply transformation on incoming tuples and may produce an output for other bolts.

\begin{figure}[ht]
\begin{center}
\begin{tikzpicture}[fig, node distance=2mm]
  \node[box] (nimbus) at (0,0) {Nimbus};
  \node[store] (zk2) at (2.6,0) {ZooKeeper};
  \node[store, above=of zk2] (zk1) {ZooKeeper};
  \node[store, below=of zk2] (zk3) {ZooKeeper};
  \node[group, fit=(zk1)(zk3), label={[title]above:ZooKeeper ensemble}] (zk) {};
  \node[box] (s2) at (5.2,0.45) {Supervisor};
  \node[box, above=of s2] (s1) {Supervisor};
  \node[box, below=of s2] (s3) {Supervisor};
  \node[box, below=of s3] (s4) {Supervisor};
  \draw[biflow] (nimbus) -- (zk.west |- nimbus);
  \foreach \s in {s1,s2,s3,s4} \draw[biflow] (zk.east |- \s) -- (\s);
\end{tikzpicture}
\caption{Storm architecture with Nimbus, ZooKeeper and supervisor nodes. Nimbus coordinates with supervisors through ZooKeeper.}
\label{fig:storm}
\end{center}
\end{figure}

Storm follows master/worker architecture, where master and worker nodes are called Nimbus and supervisors respectively.
Nimbus is responsible for distributing and coordinating the execution of topologies. 
On the other hand, supervisors are worker nodes that host one or more workers, where each worker process is mapped to a single topology.
Each worker process executes on top of a JVM and holds multiple executors.
Further, executors are made up of one or more task.
The execution of a bolt and a spout is performed in a task.
Executors provide intra-topology parallelism and task provide intra-operator parallelism.

All coordination between Nimbus and supervisors happen through ZooKeeper (see Figure \ref{fig:storm}).
Storm uses ZooKeeper to maintain its cluster state.
Nimbus is responsible for scheduling topologies on worker nodes, while the processing of each tuple is tracked by dedicated acker tasks.

Storm is fault tolerant in a way that when a worker process dies, the supervisor automatically restarts it.
In case of a node failure, the task assigned to that machine will timeout and will be assigned to another worker.
Both Nimbus and supervisor are fail-fast and stateless.
So, when Nimbus or supervisor fails, they get restarted.

Storm employs three different reliability models, i.e., at-most-once, at-least-once and exactly-once (using Trident).
However, Storm requires out-of-the-box implementation of checkpointing and recovery mechanism for stateful operators.
At-most-once reliability model is provided by basic Storm implementation.
At-least-once reliability model requires Storm topologies to keep track of every tuple in memory, while it is processed and acknowledged by downstream operator. 
In case of failure or timeout, upstream operators can replay in-memory tuples. 
Lastly, exactly-once semantics requires strict ordering on transaction, using Trident.
It uses small batches of tuples and assigns them unique transaction identifiers (TXID).
TXID is logged in external storage along with the state of the operator and is used to figure out the state of a failed tuple.
A batch requires re-submission in case of unmatched TXID.
This scheme reduces the throughput of the application. 
However, it provides strict exactly-once semantics.
\subsection{Flink}
\label{subsec:flink}
Flink is a fast and versatile framework for efficient, distributed, general-purpose data processing, which was developed at TU Berlin.
It provides the support for writing both batch and streaming applications in Java and Scala.
Flink contains custom memory management to guarantee efficient, adaptive, and highly robust switching between in-memory and data processing out-of-core algorithms.
Further, it contains its own automatic program optimizer and high-performance distributed runtime. 
It has native support for iterations, incremental iterations, and programs consisting of large DAGs of operations.

When the Flink system is started, it bring up the JobManager and one or more TaskManagers. 
The JobManager is the coordinator of the Flink system, while the TaskManagers are the workers that execute parts of the parallel programs. 
When a program is submitted, a client is created that performs the pre-processing and turns the program into the parallel data flow form that is executed by the JobManager and TaskManagers. 
Figure \ref{fig:flink} illustrates the different actors in the system very coarsely.
\begin{figure}[ht]
\begin{center}
\begin{tikzpicture}[fig]
  \node[box, fill=white] (client) at (0,2.4) {\textbf{Client}\\[-1pt]{\scriptsize compile \&}\\[-2pt]{\scriptsize optimize}};
  \node[box] (jm) at (4.0,2.4) {\textbf{JobManager}\\[-1pt]{\scriptsize scheduling,}\\[-2pt]{\scriptsize checkpoint coordination}};
  \node[box] (tm1) at (1.6,0) {\textbf{TaskManager}\\[-1pt]{\scriptsize task execution}};
  \node[box] (tm2) at (6.1,0) {\textbf{TaskManager}\\[-1pt]{\scriptsize task execution}};
  \draw[flow] (client) -- node[lbl, above] {submit\\job} (jm);
  \draw[flow] ([xshift=-9mm]jm.south) -- node[lbl, left=3pt, pos=0.5] {deploy\\tasks} ([xshift=-3mm]tm1.north);
  \draw[flow] ([xshift=4mm]tm1.north) -- node[lbl, right, pos=0.5] {status} ([xshift=-4mm]jm.south);
  \draw[biflow] ([xshift=9mm]jm.south) -- ([xshift=-4mm]tm2.north);
  \draw[biflow] (tm1) -- node[lbl, below=1mm] {data exchange} (tm2);
\end{tikzpicture}
\caption{Flink architecture with a client submitting the job to the JobManager. The JobManager schedules the job on the TaskManagers and coordinates checkpoints.}
\label{fig:flink}
\end{center}
\end{figure}

Flink provides fault tolerance for streaming operators using Asynchronous Barrier Snapshotting (ABS) \cite{carbone2015lightweight}, a variant of the Chandy-Lamport distributed snapshot algorithm.
The JobManager periodically injects barriers into the input streams at the sources.
Barriers flow through the DAG along with the regular tuples, and each operator snapshots its state when it has received the barrier from all of its inputs.
The snapshots are written asynchronously to durable storage, and together they form a globally consistent checkpoint of the application.
On failure, all operators are restored from the latest completed checkpoint and the sources, e.g., Kafka, replay the input from the positions recorded in that checkpoint.
This combination of passive replication and upstream backup at the sources provides exactly-once semantics for the operator state.

\spara{Developments since 2016.}
Flink has since become a widely used DSPE, and its state management has evolved considerably \cite{carbone2017state}.
Keyed state is partitioned into key groups, which allows the state to be redistributed when the parallelism of an operator changes.
State is kept in pluggable state backends, either on the JVM heap or in an embedded RocksDB key-value store on local disk, which supports state larger than memory and incremental checkpoints that only persist the changes since the previous checkpoint.
Besides periodic checkpoints for failure recovery, Flink offers savepoints, i.e., checkpoints that are triggered manually and used to upgrade, rescale or migrate an application without losing its state.
End-to-end exactly-once output is provided by sinks that commit their writes transactionally when a checkpoint completes.
Unaligned checkpoints\footnote{\url{https://cwiki.apache.org/confluence/display/FLINK/FLIP-76\%3A+Unaligned+Checkpoints}} allow barriers to overtake in-flight tuples, which are then stored as part of the checkpoint, to reduce checkpointing time under backpressure.
Lastly, Flink 2.0 introduces disaggregated state management, where the primary copy of the state resides in remote storage and local disks act as a cache, which substantially reduces checkpointing, recovery and rescaling times for large states \cite{mei2025disaggregated}.
\subsection{Spark Streaming}
Discretized Streams (D-Streams) \cite{zaharia2012discretized}, is a programming model for streaming applications that offers strong consistency, and efficient fault recovery.
D-Streams maps streaming applications as a series of deterministic batch of computation \cite{zaharia2012resilient}. 
It keeps the state of operators in RDD, which is a storage abstraction.
In case of failure, the system tracks the missing RDD partitions, and restores the missing state through recomputation.
This scheme recovers the state from the last checkpoint in parallel, which speeds up the recovery process. 
The major difference in D-Streams and other DSPEs is the new state management abstraction, which checkpoints group of events, rather than individual  events.

\spara{Developments since 2016.}
Structured Streaming \cite{armbrust2018structured} has since replaced D-Streams as the main streaming API of Spark.
It expresses streaming computations as incremental queries over unbounded tables and, by default, executes them as a series of micro-batches.
Before each micro-batch, Spark records the range of source offsets that it will process in a write-ahead log, and the state of stateful operators is checkpointed to durable storage after each micro-batch.
On failure, Spark restores the latest committed state and re-executes the logged offset ranges from replayable sources, e.g., Kafka.
Together with idempotent or transactional sinks, this provides end-to-end exactly-once semantics, i.e., a combination of passive replication and upstream backup at the sources.
Structured Streaming also offers a continuous processing mode with lower latency that provides at-least-once guarantees, and an embedded RocksDB-based state store\footnote{\url{https://spark.apache.org/docs/latest/streaming/apis-on-dataframes-and-datasets.html}} for large states.
\subsection{Samza}
Samza\footnote{http://samza.incubator.apache.org/} is a stream processing framework developed at LinkedIn. 
It is a distributed stream processing framework that uses Apache Kafka \cite{kreps2011kafka} (for messaging) and Apache Hadoop YARN \cite{vavilapalli2013apache} (for fault tolerance, processor isolation, security, and resource management).
Operators are represented as Samza jobs, which are responsible for applying transformations on the incoming streams.
A job is further parallelized by splitting it into multiple tasks, which are distributed across a cluster (a task in Samza is similar to a bolt in Storm).

Samza works with YARN to provide fault tolerance.
Each task includes an embedded key-value store, which is used for efficient state management.
For fault tolerance, every update to the key-value store is also written to a changelog stream in Kafka. 
In case of failure, it moves the tasks to another machine in the cluster and restores the store from the changelog.
It provides at-least-once processing and uses upstream backup techniques for handling message failures. 
\subsection{MillWheel}
MillWheel \cite{akidau2013millwheel} is a fault tolerant and scalable DSPE that was developed at Google.
In MillWheel, each event is represented using a triples \tuple{key, value, timestamp}.
In contrast to other DSPEs, it contains an additional timestamp parameter.
MillWheel uses these timestamps to compute a logical notion of time called the low watermark, a bound on the timestamps of all pending work at each stage of the pipeline.
It allows the system to handle unordered and delayed events and enables the system to easily execute time-based aggregations. 

MillWheel deduplicates events using unique record identifiers, so that their effects are applied in an idempotent manner.
It provides fault tolerance at framework level, where any component of DAG can fail without affecting the correctness of application.
MillWheel provides exactly-once semantics for event processing.
It leverages fine-grained checkpointing for the tuples, which enables efficient usage of memory at the output queues of senders.
Moreover, it provides fault tolerance at operator level using external storage.

\section{Discussion \& Conclusion}
\label{sec:discussion}
In this survey, we summarized approaches for fault tolerance in distributed stream processing systems.
There are three major fault tolerance approaches, i.e., active replication, passive replication and upstream backup, which are used in different forms.
Figure \ref{fig:tradeoff} summarizes, qualitatively, how these approaches and the combination of passive replication with upstream backup compare in terms of latency, storage cost and recovery time.

\begin{figure}[t]
\begin{center}
\definecolor{rankA}{HTML}{DCE8F6}\definecolor{rankB}{HTML}{A9C6EA}
\definecolor{rankC}{HTML}{6A9BD6}\definecolor{rankD}{HTML}{2F6DB5}
\begin{tikzpicture}[fig, x=14.5mm, y=6.5mm]
  \newcommand{\cell}[3]{
    \ifcase#3\or\def\cc{rankA}\def\tc{figink}\or\def\cc{rankB}\def\tc{figink}%
      \or\def\cc{rankC}\def\tc{white}\or\def\cc{rankD}\def\tc{white}\fi
    \node[circle, fill=\cc, draw=white, line width=1pt, minimum size=5.2mm, inner sep=0pt,
      text=\tc, font=\sffamily\footnotesize\bfseries] at (#1,-#2) {#3};}
  \foreach \c/\t in {1/Latency, 2/Storage\\cost, 3/Recovery\\time}
    \node[title, align=center, anchor=south] at (\c,-0.35) {\t};
  \foreach \r/\t in {1/Active replication (AR), 2/Passive replication (PR), 3/Upstream backup (UB), 4/PR + UB (PU)}
    \node[anchor=east, font=\sffamily\footnotesize] at (0.5,-\r) {\t};
  \foreach \r in {1,...,4} \draw[figink!15, line width=0.4pt] (0.55,-\r) -- (3.45,-\r);
  \cell{1}{1}{2}\cell{2}{1}{4}\cell{3}{1}{1}
  \cell{1}{2}{1}\cell{2}{2}{2}\cell{3}{2}{3}
  \cell{1}{3}{4}\cell{2}{3}{1}\cell{3}{3}{2}
  \cell{1}{4}{3}\cell{2}{4}{3}\cell{3}{4}{4}
  \node[lbl, anchor=north] at (2,-4.55) {rank: 1 = lowest (best) \enspace\dots\enspace 4 = highest};
\end{tikzpicture}
\caption{Qualitative comparison of active replication (AR), passive replication (PR), upstream backup (UB) and the combination of passive replication and upstream backup (PU) by their relative latency, storage cost and recovery time, from 1 (lowest) to 4 (highest). The low storage cost of UB assumes operators whose state depends on a bounded window of recent input (see Section \ref{sec:analysis}).}
\label{fig:tradeoff}
\end{center}
\end{figure}
Active replication is the most expensive of the three schemes in terms of storage.
However, it has the fastest recovery time compared to other approaches. 
Hence, active replication is an appropriate choice for time-critical DSPEs.
Active replication is capable of handling datacenter-level, node-level and task level failures.

Passive replication is less expensive as compared to active replication, in terms of storage cost and latency.
However, it requires additional state management service that can store and recover states for streaming application.
This scheme is complicated in terms of implementation.
Still, it is often used in DSPEs due to its cost effectiveness.
Further, it requires higher recovery time to recover states from the external services. 
Efficient techniques have been proposed to store states, i.e., using distributed storage or using software transactional memory.
Also, approaches have been proposed to recover states in parallel from these storage services.
Passive replication is capable of recovering node-level and task-level failures.

Upstream backup is the least expensive of the three schemes in terms of storage, for operators whose state depends on a bounded window of recent input.
However, it is the most expensive in terms of latency and throughput.
This approach requires global ordering for input stream for its implementation, e.g., Trident in Storm, low watermarks in MillWheel and sequence numbers in TimeStream.
Upstream backup is capable of handling task-level failures.

The most common approach used in DSPEs is a combination of passive replication with upstream backup, e.g., in Flink and Spark (Section \ref{sec:stream}).
In this scheme, nodes periodically take snapshots of its local state and store it in the external storage.
Moreover, upstream operators maintain the tuples after the last checkpoint in its output queue.
In case of failure, both the external storage and output queue is leveraged to recover the state of the failed operator to achieve precise recovery.
This scheme is capable of handling datacenter level, node-level and task level failures.

\section{Acknowledgment}
This work was produced during the PhD studies of the author that was supported by iSocial EU Marie Curie ITN project (FP7-PEOPLE-2012-ITN).
Also, we would like to express our gratitude to all the anonymous reviewers for their constructive feedback.

\bibliographystyle{IEEEtran}

\bibliography{biblio}

\begin{thebibliography}{10}
\providecommand{\url}[1]{#1}
\csname url@samestyle\endcsname
\providecommand{\newblock}{\relax}
\providecommand{\bibinfo}[2]{#2}
\providecommand{\BIBentrySTDinterwordspacing}{\spaceskip=0pt\relax}
\providecommand{\BIBentryALTinterwordstretchfactor}{4}
\providecommand{\BIBentryALTinterwordspacing}{\spaceskip=\fontdimen2\font plus
\BIBentryALTinterwordstretchfactor\fontdimen3\font minus
  \fontdimen4\font\relax}
\providecommand{\BIBforeignlanguage}[2]{{%
\expandafter\ifx\csname l@#1\endcsname\relax
\typeout{** WARNING: IEEEtran.bst: No hyphenation pattern has been}%
\typeout{** loaded for the language `#1'. Using the pattern for}%
\typeout{** the default language instead.}%
\else
\language=\csname l@#1\endcsname
\fi
#2}}
\providecommand{\BIBdecl}{\relax}
\BIBdecl

\bibitem{neumeyer2010s4}
L.~Neumeyer, B.~Robbins, A.~Nair, and A.~Kesari, ``S4: Distributed stream
  computing platform,'' in \emph{Data Mining Workshops (ICDMW), 2010 IEEE
  International Conference on}.\hskip 1em plus 0.5em minus 0.4em\relax IEEE,
  2010, pp. 170--177.

\bibitem{toshniwal2014storm}
A.~Toshniwal, S.~Taneja, A.~Shukla, K.~Ramasamy, J.~M. Patel, S.~Kulkarni,
  J.~Jackson, K.~Gade, M.~Fu, J.~Donham \emph{et~al.}, ``Storm@ twitter,'' in
  \emph{Proceedings of the 2014 ACM SIGMOD international conference on
  Management of data}.\hskip 1em plus 0.5em minus 0.4em\relax ACM, 2014, pp.
  147--156.

\bibitem{abadi2005design}
D.~J. Abadi, Y.~Ahmad, M.~Balazinska, U.~Cetintemel, M.~Cherniack, J.-H. Hwang,
  W.~Lindner, A.~Maskey, A.~Rasin, E.~Ryvkina \emph{et~al.}, ``The design of
  the borealis stream processing engine.'' in \emph{CIDR}, vol.~5, 2005, pp.
  277--289.

\bibitem{shah2003flux}
M.~A. Shah, J.~M. Hellerstein, S.~Chandrasekaran, and M.~J. Franklin, ``Flux:
  An adaptive partitioning operator for continuous query systems,'' in
  \emph{Data Engineering, 2003. Proceedings. 19th International Conference
  on}.\hskip 1em plus 0.5em minus 0.4em\relax IEEE, 2003, pp. 25--36.

\bibitem{zaharia2012discretized}
M.~Zaharia, T.~Das, H.~Li, S.~Shenker, and I.~Stoica, ``Discretized streams: an
  efficient and fault-tolerant model for stream processing on large clusters,''
  in \emph{Proceedings of the 4th USENIX conference on Hot Topics in Cloud
  Ccomputing}.\hskip 1em plus 0.5em minus 0.4em\relax USENIX Association, 2012,
  pp. 10--10.

\bibitem{gedik2014partitioning}
B.~Gedik, ``Partitioning functions for stateful data parallelism in stream
  processing,'' \emph{The VLDB Journal}, vol.~23, no.~4, pp. 517--539, 2014.

\bibitem{marz2015big}
N.~Marz and J.~Warren, \emph{Big Data: Principles and best practices of
  scalable realtime data systems}.\hskip 1em plus 0.5em minus 0.4em\relax
  Manning Publications Co., 2015.

\bibitem{stonebraker20058}
M.~Stonebraker, U.~{\c{C}}etintemel, and S.~Zdonik, ``The 8 requirements of
  real-time stream processing,'' \emph{ACM SIGMOD Record}, vol.~34, no.~4, pp.
  42--47, 2005.

\bibitem{zaharia2012resilient}
M.~Zaharia, M.~Chowdhury, T.~Das, A.~Dave, J.~Ma, M.~McCauley, M.~J. Franklin,
  S.~Shenker, and I.~Stoica, ``Resilient distributed datasets: A fault-tolerant
  abstraction for in-memory cluster computing,'' in \emph{Proceedings of the
  9th USENIX conference on Networked Systems Design and Implementation}.\hskip
  1em plus 0.5em minus 0.4em\relax USENIX Association, 2012, pp. 2--2.

\bibitem{hindman2011mesos}
B.~Hindman, A.~Konwinski, M.~Zaharia, A.~Ghodsi, A.~D. Joseph, R.~H. Katz,
  S.~Shenker, and I.~Stoica, ``Mesos: A platform for fine-grained resource
  sharing in the data center.'' in \emph{NSDI}, vol.~11, 2011, pp. 22--22.

\bibitem{vavilapalli2013apache}
V.~K. Vavilapalli, A.~C. Murthy, C.~Douglas, S.~Agarwal, M.~Konar, R.~Evans,
  T.~Graves, J.~Lowe, H.~Shah, S.~Seth \emph{et~al.}, ``Apache hadoop yarn: Yet
  another resource negotiator,'' in \emph{Proceedings of the 4th annual
  Symposium on Cloud Computing}.\hskip 1em plus 0.5em minus 0.4em\relax ACM,
  2013, p.~5.

\bibitem{dean2008mapreduce}
J.~Dean and S.~Ghemawat, ``Mapreduce: simplified data processing on large
  clusters,'' \emph{Communications of the ACM}, vol.~51, no.~1, pp. 107--113,
  2008.

\bibitem{condie2010mapreduce}
T.~Condie, N.~Conway, P.~Alvaro, J.~M. Hellerstein, K.~Elmeleegy, and R.~Sears,
  ``Mapreduce online.'' in \emph{NSDI}, vol.~10, 2010, p.~20.

\bibitem{murphy2006naive}
K.~P. Murphy, ``Naive bayes classifiers,'' \emph{University of British
  Columbia}, 2006.

\bibitem{shah2004highly}
M.~A. Shah, J.~M. Hellerstein, and E.~Brewer, ``Highly available,
  fault-tolerant, parallel dataflows,'' in \emph{Proceedings of the 2004 ACM
  SIGMOD international conference on Management of data}.\hskip 1em plus 0.5em
  minus 0.4em\relax ACM, 2004, pp. 827--838.

\bibitem{balazinska2008fault}
M.~Balazinska, H.~Balakrishnan, S.~R. Madden, and M.~Stonebraker,
  ``Fault-tolerance in the borealis distributed stream processing system,''
  \emph{ACM Transactions on Database Systems (TODS)}, vol.~33, no.~1, p.~3,
  2008.

\bibitem{ananthanarayanan2013photon}
R.~Ananthanarayanan, V.~Basker, S.~Das, A.~Gupta, H.~Jiang, T.~Qiu,
  A.~Reznichenko, D.~Ryabkov, M.~Singh, and S.~Venkataraman, ``Photon:
  Fault-tolerant and scalable joining of continuous data streams,'' in
  \emph{Proceedings of the 2013 international conference on Management of
  data}.\hskip 1em plus 0.5em minus 0.4em\relax ACM, 2013, pp. 577--588.

\bibitem{loesing2012stormy}
S.~Loesing, M.~Hentschel, T.~Kraska, and D.~Kossmann, ``Stormy: an elastic and
  highly available streaming service in the cloud,'' in \emph{Proceedings of
  the 2012 Joint EDBT/ICDT Workshops}.\hskip 1em plus 0.5em minus 0.4em\relax
  ACM, 2012, pp. 55--60.

\bibitem{castro2013integrating}
R.~Castro~Fernandez, M.~Migliavacca, E.~Kalyvianaki, and P.~Pietzuch,
  ``Integrating scale out and fault tolerance in stream processing using
  operator state management,'' in \emph{Proceedings of the 2013 international
  conference on Management of data}.\hskip 1em plus 0.5em minus 0.4em\relax
  ACM, 2013, pp. 725--736.

\bibitem{kwon2008fault}
Y.~Kwon, M.~Balazinska, and A.~Greenberg, ``Fault-tolerant stream processing
  using a distributed, replicated file system,'' \emph{Proceedings of the VLDB
  Endowment}, vol.~1, no.~1, pp. 574--585, 2008.

\bibitem{brito2009minimizing}
A.~Brito, C.~Fetzer, and P.~Felber, ``Minimizing latency in fault-tolerant
  distributed stream processing systems,'' in \emph{Distributed Computing
  Systems, 2009. ICDCS'09. 29th IEEE International Conference on}.\hskip 1em
  plus 0.5em minus 0.4em\relax IEEE, 2009, pp. 173--182.

\bibitem{sebepou2011cec}
Z.~Sebepou and K.~Magoutis, ``Cec: Continuous eventual checkpointing for data
  stream processing operators,'' in \emph{Dependable Systems \& Networks (DSN),
  2011 IEEE/IFIP 41st International Conference on}.\hskip 1em plus 0.5em minus
  0.4em\relax IEEE, 2011, pp. 145--156.

\bibitem{zhou2010programming}
J.~Zhou, C.~Zhang, H.~Tang, J.~Wu, and T.~Yang, ``Programming support and
  adaptive checkpointing for high-throughput data services with log-based
  recovery,'' in \emph{Dependable Systems and Networks (DSN), 2010 IEEE/IFIP
  International Conference on}.\hskip 1em plus 0.5em minus 0.4em\relax IEEE,
  2010, pp. 91--100.

\bibitem{qian2013timestream}
Z.~Qian, Y.~He, C.~Su, Z.~Wu, H.~Zhu, T.~Zhang, L.~Zhou, Y.~Yu, and Z.~Zhang,
  ``Timestream: Reliable stream computation in the cloud,'' in
  \emph{Proceedings of the 8th ACM European Conference on Computer
  Systems}.\hskip 1em plus 0.5em minus 0.4em\relax ACM, 2013, pp. 1--14.

\bibitem{akidau2013millwheel}
T.~Akidau, A.~Balikov, K.~Bekiro{\u{g}}lu, S.~Chernyak, J.~Haberman, R.~Lax,
  S.~McVeety, D.~Mills, P.~Nordstrom, and S.~Whittle, ``Millwheel:
  fault-tolerant stream processing at internet scale,'' \emph{Proceedings of
  the VLDB Endowment}, vol.~6, no.~11, pp. 1033--1044, 2013.

\bibitem{hwang2005high}
J.-H. Hwang, M.~Balazinska, A.~Rasin, U.~Cetintemel, M.~Stonebraker, and
  S.~Zdonik, ``High-availability algorithms for distributed stream
  processing,'' in \emph{Data Engineering, 2005. ICDE 2005. Proceedings. 21st
  International Conference on}.\hskip 1em plus 0.5em minus 0.4em\relax IEEE,
  2005, pp. 779--790.

\bibitem{cristian1991understanding}
F.~Cristian, ``Understanding fault-tolerant distributed systems,''
  \emph{Communications of the ACM}, vol.~34, no.~2, pp. 56--78, 1991.

\bibitem{gray1996dangers}
J.~Gray, P.~Helland, P.~O'Neil, and D.~Shasha, ``The dangers of replication and
  a solution,'' in \emph{ACM SIGMOD Record}, vol.~25, no.~2.\hskip 1em plus
  0.5em minus 0.4em\relax ACM, 1996, pp. 173--182.

\bibitem{brito2008speculative}
A.~Brito, C.~Fetzer, H.~Sturzrehm, and P.~Felber, ``Speculative out-of-order
  event processing with software transaction memory,'' in \emph{Proceedings of
  the second international conference on Distributed event-based
  systems}.\hskip 1em plus 0.5em minus 0.4em\relax ACM, 2008, pp. 265--275.

\bibitem{7113328}
Y.~Wu and K.-L. Tan, ``Chronostream: Elastic stateful stream computation in the
  cloud,'' in \emph{Data Engineering (ICDE), 2015 IEEE 31st International
  Conference on}, April 2015, pp. 723--734.

\bibitem{ali2009microsoft}
M.~H. Ali, C.~Gerea, B.~S. Raman, B.~Sezgin, T.~Tarnavski, T.~Verona, P.~Wang,
  P.~Zabback, A.~Ananthanarayan, A.~Kirilov \emph{et~al.}, ``Microsoft cep
  server and online behavioral targeting,'' \emph{Proceedings of the VLDB
  Endowment}, vol.~2, no.~2, pp. 1558--1561, 2009.

\bibitem{boykin2014summingbird}
O.~Boykin, S.~Ritchie, I.~O’Connell, and J.~Lin, ``Summingbird: A framework
  for integrating batch and online mapreduce computations,'' \emph{Proceedings
  of the VLDB Endowment}, vol.~7, no.~13, 2014.

\bibitem{agha1985actors}
G.~A. Agha, ``Actors: a model of concurrent computation in distributed
  systems,'' 1985.

\bibitem{zookeeper}
A.~S. Foundation, ``Apache zookeeper,'' "\url{http://zookeeper.apache.org/}".

\bibitem{kreps2011kafka}
J.~Kreps, N.~Narkhede, J.~Rao \emph{et~al.}, ``Kafka: A distributed messaging
  system for log processing,'' in \emph{Proceedings of the NetDB}, 2011.

\bibitem{carbone2015lightweight}
P.~Carbone, G.~F{\'o}ra, S.~Ewen, S.~Haridi, and K.~Tzoumas, ``Lightweight
  asynchronous snapshots for distributed dataflows,'' \emph{arXiv preprint
  arXiv:1506.08603}, 2015.

\bibitem{to2018survey}
Q.-C. To, J.~Soto, and V.~Markl, ``A survey of state management in big data
  processing systems,'' \emph{The VLDB Journal}, vol.~27, no.~6, pp.
  847--872, 2018.

\bibitem{wang2022comprehensive}
X.~Wang, C.~Zhang, J.~Fang, R.~Zhang, W.~Qian, and A.~Zhou, ``A comprehensive
  study on fault tolerance in stream processing systems,'' \emph{Frontiers of
  Computer Science}, vol.~16, no.~2, p. 162603, 2022.

\bibitem{fragkoulis2024survey}
M.~Fragkoulis, P.~Carbone, V.~Kalavri, and A.~Katsifodimos, ``A survey on the
  evolution of stream processing systems,'' \emph{The VLDB Journal}, vol.~33,
  no.~2, pp. 507--541, 2024.

\bibitem{carbone2017state}
P.~Carbone, S.~Ewen, G.~F{\'o}ra, S.~Haridi, S.~Richter, and K.~Tzoumas,
  ``State management in {Apache Flink}: Consistent stateful distributed stream
  processing,'' \emph{Proceedings of the VLDB Endowment}, vol.~10, no.~12, pp.
  1718--1729, 2017.

\bibitem{mei2025disaggregated}
Y.~Mei, R.~Xia, Z.~Lan, K.~Hu, L.~Huang, P.~Carbone, Y.~Lei, V.~Kalavri,
  H.~Yin, and F.~Wang, ``Disaggregated state management in {Apache
  Flink} 2.0,'' \emph{Proceedings of the VLDB Endowment}, vol.~18, no.~12, pp.
  4846--4859, 2025.

\bibitem{armbrust2018structured}
M.~Armbrust, T.~Das, J.~Torres, B.~Yavuz, S.~Zhu, R.~Xin, A.~Ghodsi,
  I.~Stoica, and M.~Zaharia, ``Structured streaming: A declarative {API} for
  real-time applications in {Apache Spark},'' in \emph{Proceedings of the 2018
  International Conference on Management of Data (SIGMOD)}.\hskip 1em plus
  0.5em minus 0.4em\relax ACM, 2018, pp. 601--613.

\end{thebibliography}

\end{document}